\documentclass[letterpaper]{article}

\usepackage{aaai20,times,helvet,courier,graphicx}
\usepackage[hyphens]{url}
\usepackage{amsfonts,microtype,booktabs,tabularx,glossaries,subcaption}

\title{Perceiving Music Quality with GANs}
\author{Agrin Hilmkil, Carl Thom\'e, Anders Arpteg\\ Peltarion}

\newcolumntype{R}{>{\small\raggedleft\arraybackslash}X}

\newacronym{qa}{QA}{quality assessment}
\newacronym{mos}{MOS}{mean opinion score}
\newacronym{mse}{MSE}{mean squared error}
\newacronym{gan}{GAN}{generative adversarial network}
\newacronym{stft}{STFT}{short-time Fourier transform}
\newacronym{is}{IS}{Inception score}
\newacronym{fid}{FID}{Fr\'echet Inception distance}
\newacronym{daw}{DAW}{digital audio workstation}
\newacronym{dbn}{DBN}{deep belief network}
\newacronym{ann}{ANN}{artifical neural network}
\newacronym{amt}{AMT}{Amazon Mechanical Turk}
\newacronym{sf}{SF}{spectral flatness}

\begin{document}
\maketitle
\begin{abstract}
Several methods have been developed to assess the perceptual quality of audio under transforms like lossy compression. However, they require paired reference signals of the unaltered content, limiting their use in applications where references are unavailable. This has hindered progress in audio generation and style transfer, where a no-reference quality assessment method would allow more reproducible comparisons across methods. We propose training a GAN on a large music library, and using its discriminator as a no-reference quality assessment measure of the perceived quality of music. This method is unsupervised, needs no access to degraded material and can be tuned for various domains of music. In a listening test with 448 human subjects, where participants rated professionally produced music tracks degraded with different levels and types of signal degradations such as waveshaping distortion and low-pass filtering, we establish a dataset of human rated material. By using the human rated dataset we show that the discriminator score correlates significantly with the subjective ratings, suggesting that the proposed method can be used to create a no-reference musical audio quality assessment measure.
\end{abstract}

\section{Introduction}
\begin{figure}[ht]
\includegraphics[width=\columnwidth]{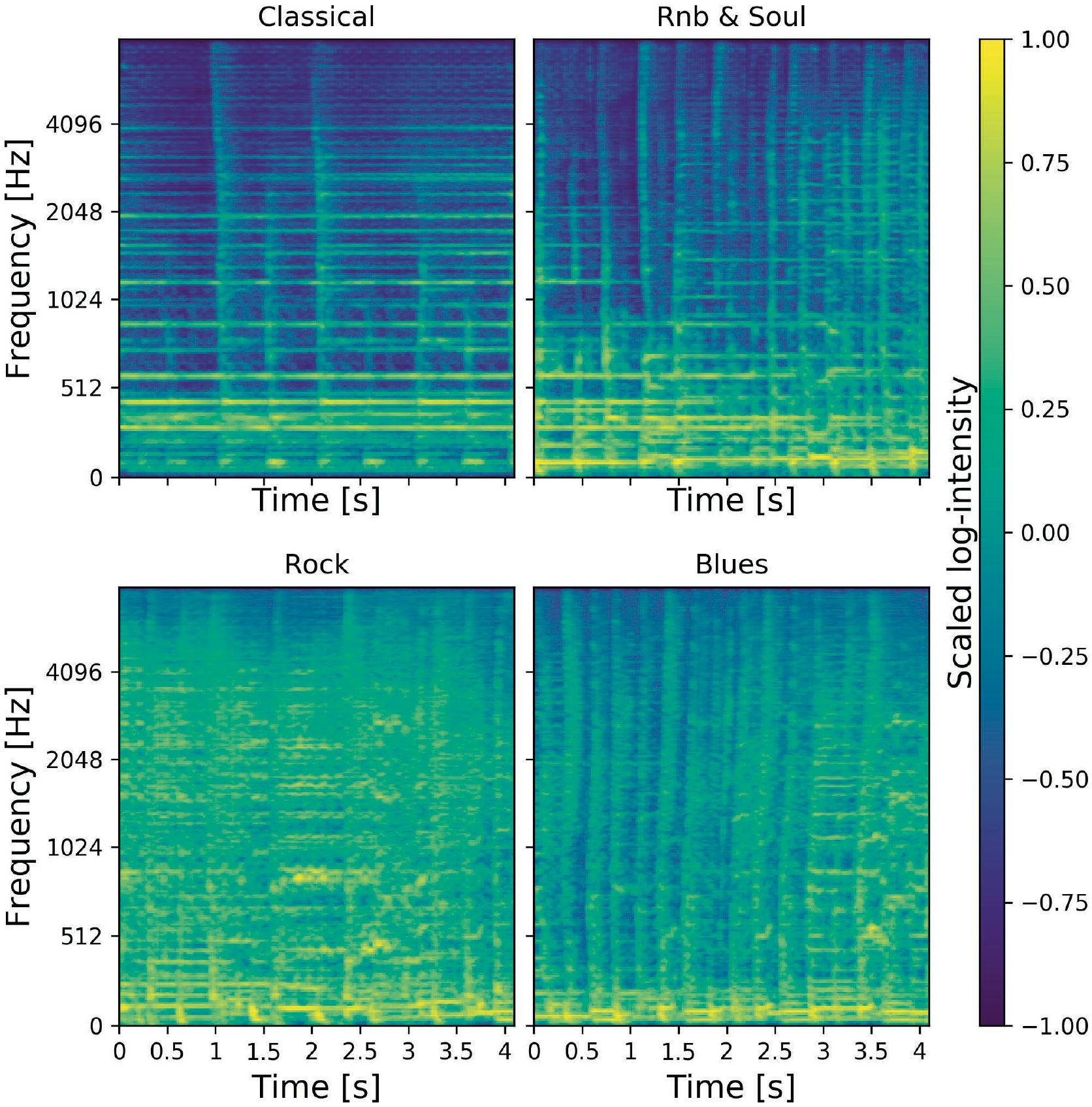}
\caption{\label{fig:generated_mel_spectrograms}A random sample of mel spectrograms produced by the generator for a random selection of genres. One may note the strong harmonics of the \textit{Classical} segment and contrast them with the more distorted ones appearing in \textit{Rock}.}
\end{figure}
Audio quality is usually estimated by the difference between a clean reference signal $y$ and a listenable output $\hat{y}$~\cite{campbell2009audio}. By computing the error signal with a metric $d(y, \hat{y})$, such as the \gls{mse} of spectrograms~\cite{hines2015}, a subsequent quality assessment can be realized by interpreting the remaining frequency content~\cite{thiede2000}.

While straightforward in principle, the \gls{mse} does not follow human perception~\cite{hines2015} and it is easy to construct signal pairs with a large distance that sound very similar, or signals that have a small distance yet sound nothing alike. Therefore considerable work has been done on developing perceptually oriented metrics such as PEAQ~\cite{thiede2000}, Barbedo's model~\cite{barbedo2005new}, Moore's model~\cite{moore2004} and PEMO-Q~\cite{huber2006}. The essence is typically to filter the signals in various ways to measure different frequencies differently, inspired by knowledge of the human auditory system and music cognition~\cite{campbell2009audio}. Aside from depending on extensive domain expertise, these methods all have a hard requirement on the existence of a reference signal, which severely limits their applicability.

For example, in order to rank live recordings~\cite{li2013} in terms of audio quality, we would first have to time align the recordings with the studio recording which is a hard problem in itself. Similarly problematic are the cases where a reference signal is unavailable or does not exist. When developing style transfer algorithms for musical audio for example~\cite{huang2018timbretron}, the current best practice for evaluating algorithms is to manually listen to sample outputs and collect a \gls{mos}. This process is typically labor intensive yet hard to replace.

The same fundamental problem hinders musical audio generation research~\cite{NIPS2018_8023} and even though rough heuristics like estimating musical consonance by detecting pitch events could potentially provide some guidance, it is probable that even the most well considered heuristic would still fail to capture important nuances like whether a physically plausible timbre is present, or whether the piece is performed with an appropriately expressive musical performance.

In short, there is simply no good method available for determining the quality of musical audio and it is crucially missing for large scale \gls{qa}. Instead of relying on subjective listening tests, employing a no-reference \gls{qa} measure would speed up algorithm development and promote reproducible research.

In this work we approach no-reference \gls{qa} by way of generative modelling. \Glspl{gan} have recently shown promise in modelling musical audio~\cite{donahue2018adversarial,engel2018gansynth}. \citeauthor{arjovsky2017wasserstein} \shortcite{arjovsky2017wasserstein} observed in their work that the perceived quality of generated content in a \gls{gan} correlates with the loss of the critic. Therefore, we propose a \gls{gan} based method for no-reference \gls{qa}. The \gls{gan} is trained to model the empirical distribution of music with high production quality. Its discriminator, adversarially trained to detect out of distribution samples, is used as a measure of perceived quality. This method is unsupervised and has the advantage of being tunable to different domains, such as genres, in order to handle interactions between genre and perception of audio effects. By establishing human quality ratings on a varied test set of both clean and degraded music signals we show that the discriminator score correlates significantly with the subjective ratings, and therefore is predictive of human perception of quality. We further validate the plausibility of our model by studying the content it generates (Figure \ref{fig:generated_mel_spectrograms}). To promote future work into this area we are releasing an open source implementation and trained model upon publication.

\section{Related work}
Historically there has been great interest in comparing and developing audio compression methods for signal transmission, leading to progress in reference based methods~\cite{campbell2009audio}. The no-reference setting has not received as much attention. Meanwhile, for image content, blind quality estimation has progressed with \glspl{dbn} even outperforming state-of-the-art reference based methods~\cite{tang2014blind}. Their approach relied mostly on unsupervised pre-training, though, as opposed to our method also applied supervised fine-tuning.

Fully supervised, discriminative models have been applied to audio, although on small datasets. \Glspl{ann} have been used to map values of perceptually-inspired features to a subjective scale of perceived quality~\cite{manders2012}. Training data consisted of values of perceptual measures obtained from ten different excerpts of orchestral music processed by a simplified model of a hearing aid with an adaptive feedback canceller, and corresponding subjective quality ratings from 27 normal hearing subjects. Another study found that quality measures employing valid auditory models generalized best across different distortions~\cite{harlander2014sound}. Their models were able to predict a large range of different distortions and performed best compared to other state-of-the-art quality measures.

A third approach is to consider quality as a ranking problem. By relating multiple versions of the same song like various recordings of the same live performance it is possible to retrieve the best sounding versions~\cite{li2013,jingli2015music}. The requirement that signals be grouped is, however, too limiting for general use.

An alternative problem formulation is to predict underlying mix settings by obtaining cues about the signal mixing process and ideally recovering exact audio effect settings~\cite{fourer2017}. Such cues need to be mapped to perceived quality, however, and that is a big task in itself. While there are reasonably objective qualities for speech (e.g. intelligibility), for musical audio what qualities are important can be ambiguous and highly context-sensitive. For example, distortion is expected in rock music and frowned upon in classical music.

\section{Method}
\subsection{Data}
The underlying dataset used is from an online service for music\footnote{\url{https://www.epidemicsound.com/music/}} of professionally produced, high-quality music. It contains a wide range of music and is curated to conform well to contemporary music, as it is intended for use by content creators. Using their catalog we created a balanced subset (Table \ref{table:data_summary}) of mutually exclusive genres.

\begin{table}[ht]
\begin{tabularx}{\linewidth}{XRRR}
\toprule
Genres & Tracks & Duration [H:M:S] & Ratio \\
\midrule
Acoustic    &     165 &  6:08:08 &      7.09\% \\
Blues       &     165 &  7:29:28 &      8.65\% \\
Classical   &     165 &  5:37:29 &      6.50\% \\
Country     &     165 &  7:02:08 &      8.13\% \\
E. \& D.    &     165 &  7:16:33 &      8.41\% \\
Funk        &     165 &  5:11:34 &      6.00\% \\
Hip-hop     &     165 &  6:57:50 &      8.05\% \\
Jazz        &     165 &  6:53:18 &      7.96\% \\
Latin       &     165 &  6:09:53 &      7.12\% \\
Pop         &     165 &  7:58:05 &      9.21\% \\
Reggae      &     165 &  5:56:39 &      6.87\% \\
Soul        &     165 &  8:01:44 &      9.28\% \\
Rock        &     165 &  5:50:31 &      6.75\% \\
\midrule
\textbf{All}&    2145 & 86:41:06 &       100\% \\
\bottomrule
\end{tabularx}
\caption{\label{table:data_summary}Summary of the size of the available dataset. \textit{Electronica \& Dance} has been abbreviated \textit{E. \& D.}. Ratio shows the ratio of genre duration to total duration.}
\end{table}

Although there is some variation in the number of hours available per genre, our training procedure further balances this data and ensures that we consume an equal amount of data from each genre. We create a training set (80\%), a test set for human evaluation (3\%) and a reserved set for future use (17\%) by uniformly sampling the proportions from each genre. All tracks are 48 kHz / 24-bit PCM stereo mixes.

\subsection{Degrading audio quality}
To include tracks of varying quality we introduce a set of signal degradations with the following open-source REAPER JSFX audio plugins~\cite{reaper_jsfx}:

\begin{itemize}
\item \textbf{Distortion} (\texttt{loser/waveShapingDstr}) Waveshaping distortion with the waveshape going from a sine-like shape (50\%) to square (100\%).
\item \textbf{Lowpass} (\texttt{Liteon/butterworth24db}) Low-pass filtering, a 24 dB Butterworth filter configured to have a frequency cutoff from 20 kHz down to 1000 Hz.
\item \textbf{Limiter} (\texttt{loser/MGA\_JSLimiter}) Mastering limiter, having all settings fixed except for the threshold that was lowered from 0 dB to -30 dB (introduces clipping artifacts).
\item \textbf{Noise} (\texttt{Liteon/pinknoisegen}) Additive noise on a range from -25 dB (subtly audible) to 0.0 dB (clearly audible).
\end{itemize}

Plugins were applied separately to each track without effects chaining. The parameter of each plugin is rescaled to $\left[0, 100\right]$ and considered the intensity of the degradation. Each time a degradation is applied an intensity is randomly chosen from the uniformly distribution of the range.

\subsection{Human perceived listening quality}
We create a dataset with music segments and their corresponding human-perceived listening quality from the test set, to evaluate our methods effectiveness. This evaluation dataset is made freely available\footnote{\url{https://github.com/Peltarion/pmqd}}. As a convenient method for getting human ratings from a wide population we turn to crowdsourcing the task on \gls{amt}. This has the advantage of allowing significantly larger scale than controlled tests, though introduces some potential problems such as cheating and underperforming participants, which we handle as described in this section.

\textbf{Music segments}\quad From the tracks in the test set we randomly pick 3 segments per track with a duration of 4 seconds, producing 195 segments. Additional segments of varying quality are created by degrading each original segment once with each degradation type, yielding 975 segments in total.

\textbf{Task assignment}\quad Tasks to be completed by human participants are created to rate segments for their listening quality. Segments are randomly assigned to tasks such that each task contains 10 segments, never contains duplicates and each segment occurs in at least 5 tasks. Participants may only perform one task, in order to avoid individuals biases. In total we produce 488 tasks resulting in 4880 individual segment evaluations.

\textbf{Task specification}\quad During a task, each participant is asked to specify which type of device they will use for listening from the list: ``smartphone speaker'', ``speaker'', ``headphones'', ``other'', ``will not listen''. If any other option than ``speaker'' or ``headphones'' is selected that submission is rejected and the task re-assigned. For each segment in the task we ask the user for an assessment of audio quality, not musical content~\cite{wilson2016}. The question is phrased as: ``How do you rate the audio quality of this music segment?'', and may be answered on the ordinal scale: ``Bad'', ``Poor'', ``Fair'', ``Good'' and ``Excellent'', corresponding to numerical values 1-5.

\textbf{Rating aggregation}\quad Once all tasks are completed the ratings are aggregated to produce one perceived quality rating per segment. Since participants are listening in their own respective environments we are concerned with lo-fi audio equipment, or scripted responses trying to game \gls{amt}. Thus we use the median over the mean rating to discount outliers.

\textbf{Cheating}\quad The following schemes are applied in an attempt to reduce cheating or participants not following instructions:
\begin{itemize}
    \item Multiple submissions by the same participant despite warnings that this will lead to rejection are all rejected
    \item Tasks completed in a shorter amount of time than the total duration of all segments in the task are rejected
    \item Tasks where all segments are given the same rating despite large variation in degradation intensity are rejected
    \item The number of tasks available at any moment is restricted to 50, as a smaller amount has been shown to decrease the prevalence of cheating~\cite{Eickhoff2013}
\end{itemize}

\subsection{Music representation}
All tracks are downsampled to mono mixes at 16 kHz / 16-bit for training the GAN. This limits the highest possible fidelity, but makes it easier to cover longer time-spans while reducing data loading time and memory-footprint.

Like SpecGAN~\cite{donahue2018adversarial} and GANSynth~\cite{engel2018gansynth} we use a time-frequency representation. This allows us to adopt existing \gls{gan} architectures, which is especially important as \glspl{gan} are notoriously difficult to train and small changes of hyperparameters often result in various issues.

Spectrograms are produced by the \gls{stft} with 2048 sample-length Hann windows, 256 samples apart. A Mel filterbank of 256 bands was applied to the magnitudes in the frequency domain to reduce the dimensionality while preserving frequency resolution in the middle register. The resulting Mel filtered spectrograms were log scaled and individually rescaled to the range $[-1, 1]$.

\subsection{Model}
Despite the similarities between images traditionally modelled by \glspl{gan} and the mel spectrograms we aim to model, there are certain differences that may make modelling harder. In particular, components of individual audio objects tend to be non-local~\cite{wyse2017spectrograms}, which may be difficult for purely convolutional models to capture due to their limited receptive field. For this reason we use the SA\gls{gan}~\cite{pmlr-v97-zhang19d}, which incorporates self-attention to compute features from the full spatial extent of a layer of activation maps. Furthermore, we maintain SA\glspl{gan} use of the projected c\gls{gan} \cite{miyato2018cgans} to allow the model to be tuned individually for the different genres, in line with the expectation that for example distortion may be expected in rock music yet be perceived as of low quality in classical music.

In our case we aim to model the distribution $p_{\mathcal{X}_y}$ of mel spectrograms $x \in \mathcal{X}_y$ from genre $y \in \mathcal{Y}$. The generator $G$ learns a mapping such that $G(z, y) \sim p_{\mathcal{X}_y}$ when $z \sim p_\mathcal{Z}$, by competing against a discriminator $D$ attempting to tell real samples $x$ apart from generated ones $G(z, y)$. The samples $z \in \mathcal{Z}$ are referred to as noise and the dimensionality of $\mathcal{Z}$ and family of $p_\mathcal{Z}$ are considered hyperparameters. Like \citeauthor{pmlr-v97-zhang19d} \shortcite{pmlr-v97-zhang19d} we alternate by minimizing the hinge-losses~\cite{lim2017geometric} corresponding to $D$ and $G$:

\begin{align}
    L_D &= \, - \mathbb{E}_{x \sim p_\mathcal{X}}\left[ \min(0, - 1 + D(x, y)) \right] \nonumber \\
               & - \mathbb{E}_{z \sim p_\mathcal{Z}}\left[\min(0, - 1 - D(G(z, y), y)\right] , \\
    L_G &= \, -\mathbb{E}_{z \sim p_\mathcal{Z}}\left[D(G(z, y), y\right] .
\end{align}

\subsection{Training parameters}
The \gls{gan} architecture used is the $256 \times 256$ Big\gls{gan}~\cite{brock2018large} but without applying any of the additional losses, and handling $z$, $y$ like in SA\gls{gan}. We set the channel width multiplier to $64$ for both generator and discriminator. The input noise $z$ is $120$ dimensions sampled from a standard normal distribution, $\mathcal{N}(0, I)$. Training is towered across 4 Titan X Pascal GPUs with 12GB memory each, which restricted our batch size to 6 samples per tower. The generator is trained with the learning rate $1\cdot 10^{-4}$ and the discriminator with the learning rate $2\cdot 10^{-4}$. Updates are done sequentially with the discriminator being updated twice for each generator step. Real samples are fed to the discriminator by randomly sampling tracks from the training set without replacement, from which a uniformly random segment is selected to construct each batch. Once all tracks have been sampled we consider an epoch to have passed and restart to produce the next epoch. Training is stopped when a batch of mel spectrograms are generated which look close to real mel spectrograms. While there are starting to appear methods for determining when to stop training of GANs we are not familiar with any that are shown to consistently work well when generating audio.

\subsection{Perceptual scoring}
In this work we refer to $D(x, y)$ as the discriminator score. When the discriminator score is correlated to human perceived music quality it is given a mel spectrogram of the same segment of audio as was rated by human annotators, but like the training data it is downsampled to 16kHz / 16bit mono mix. Furthermore, the discriminator is provided with the genre of each sample. This conveniently allows us to handle the genre dependent qualities.

\section{Results}
\subsection{Correlation with human opinion}
\begin{figure}[ht]
    \centering
    \includegraphics[width=.99\columnwidth]{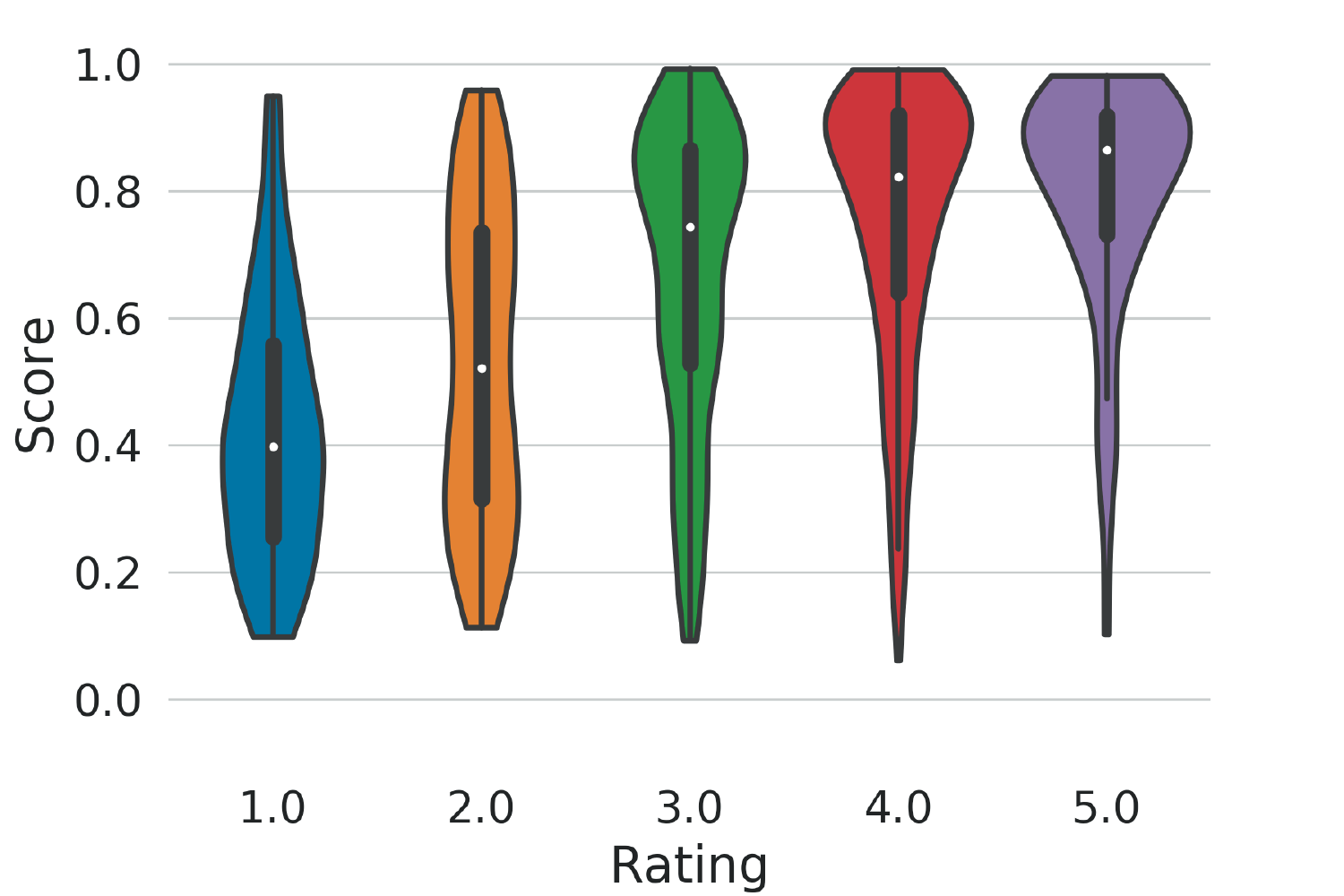}
    \caption{Violin plot illustrating the distribution of discriminator score for the median human rating of each clip, with densities truncated to the observed range of data.}
    \label{fig:score}
\end{figure}
We illustrate the distribution of the discriminator score (\textit{D}) for different ratings in Figure \ref{fig:score}. This shows that the median values of \textit{D} for each rating increase monotonically, suggesting that the method may be particularly suitable for ranking collections of data. Similar to \citeauthor{tang2014blind}~\shortcite{tang2014blind} we study the Spearman correlation between our method and the collected ratings of clips to evaluate the effectiveness of our method (Table \ref{tab:significance}). This shows that our method correlates in rank to the human rating with a high significance ($p = 3.225\cdot 10^{-44}$) over the entire rated dataset. Broken down by different subsets it is seen to perform significantly better on the genres Funk, Pop and Country. Furthermore, it shows less significant correlation with the human rating when degraded by a limiter or a low-pass filter.
\begin{table}[ht]
    \centering
    \begin{tabularx}{\linewidth}{XRX}
        \toprule
                       Subset &  Correlation &               p-value \\
        \midrule
               \textbf{Genre} & & \\
                     Acoustic &        0.426 &  $1.400\cdot 10^{-4}$ \\
                        Blues &        0.450 &  $5.657\cdot 10^{-5}$ \\
                    Classical &        0.259 &  $2.468\cdot 10^{-2}$ \\
                      Country &        0.567 &  $1.110\cdot 10^{-7}$ \\
                     E. \& D. &        0.221 &  $5.662\cdot 10^{-2}$ \\
                         Funk &        0.677 & $2.664\cdot 10^{-11}$ \\
                      Hip Hop &        0.516 &  $2.123\cdot 10^{-6}$ \\
                         Jazz &        0.469 &  $2.166\cdot 10^{-5}$ \\
                        Latin &        0.488 &  $1.031\cdot 10^{-5}$ \\
                          Pop &        0.591 &  $2.321\cdot 10^{-8}$ \\
                       Reggae &        0.520 &  $1.752\cdot 10^{-6}$ \\
                  Rnb \& Soul &        0.519 &  $1.871\cdot 10^{-6}$ \\
                         Rock &        0.304 &  $7.909\cdot 10^{-3}$ \\
        \midrule
         \textbf{Degradation} & & \\
                   Distortion &        0.349 &  $5.937\cdot 10^{-7}$ \\
                      Limiter &        0.120 &  $9.380\cdot 10^{-2}$ \\
                      Lowpass &        0.222 &  $1.830\cdot 10^{-3}$ \\
                        Noise &        0.359 &  $2.638\cdot 10^{-7}$ \\
        \midrule
                 \textbf{All} &        0.426 & $3.225\cdot 10^{-44}$ \\
        \bottomrule
    \end{tabularx}
    \caption{Spearman correlation between discriminator score and median rating with significance values for different subsets across genres, types of degradation and all data.}
    \label{tab:significance}
\end{table}

\subsection{Comparison to other measures}
\label{sec:results:comparision}
\begin{figure*}[ht]
\centering
    \begin{subfigure}[b]{.32\linewidth}
        \includegraphics[width=\linewidth]{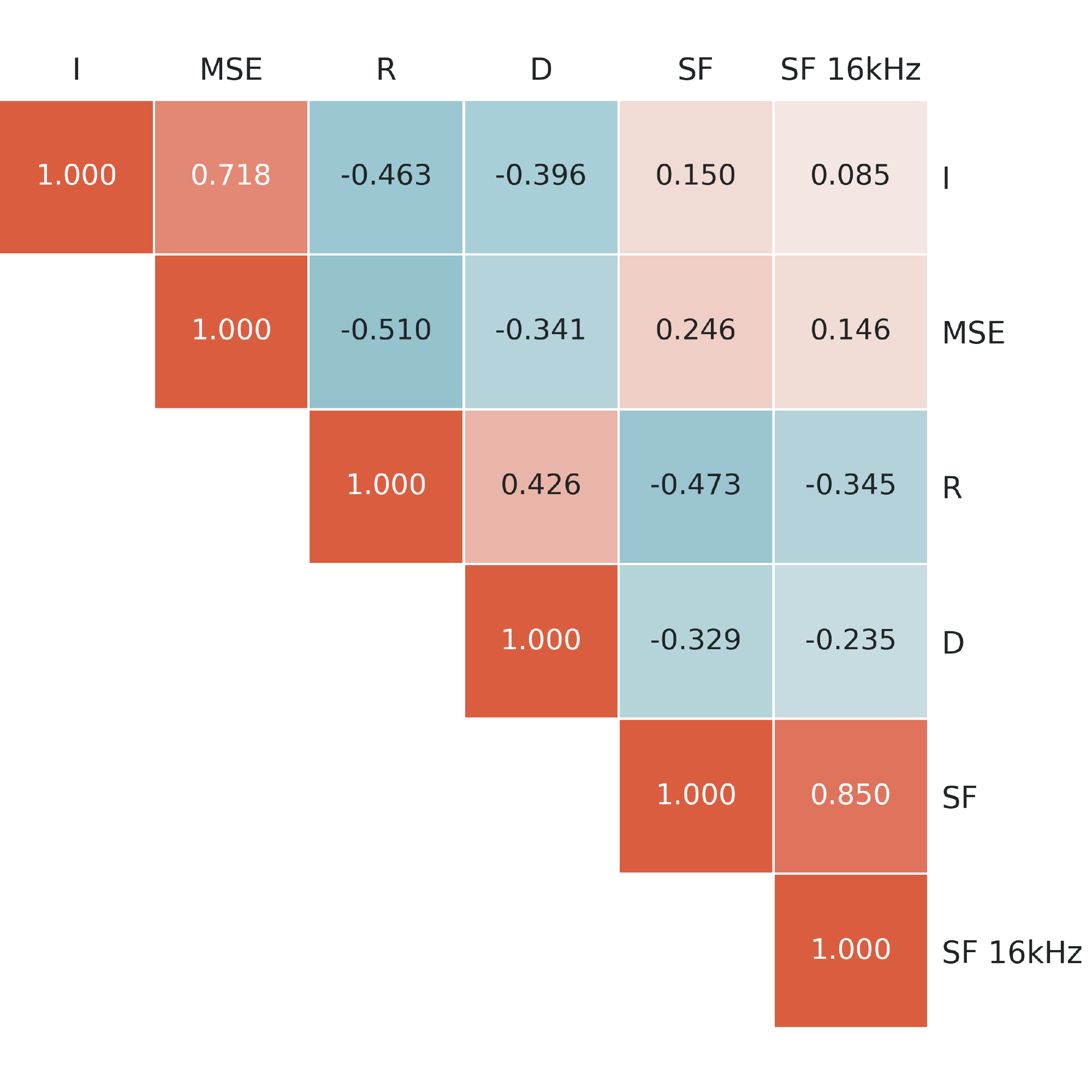}
        \subcaption{\label{fig:correlations:pairwise}}
    \end{subfigure}
    \begin{subfigure}[b]{.32\linewidth}
        \includegraphics[width=\linewidth]{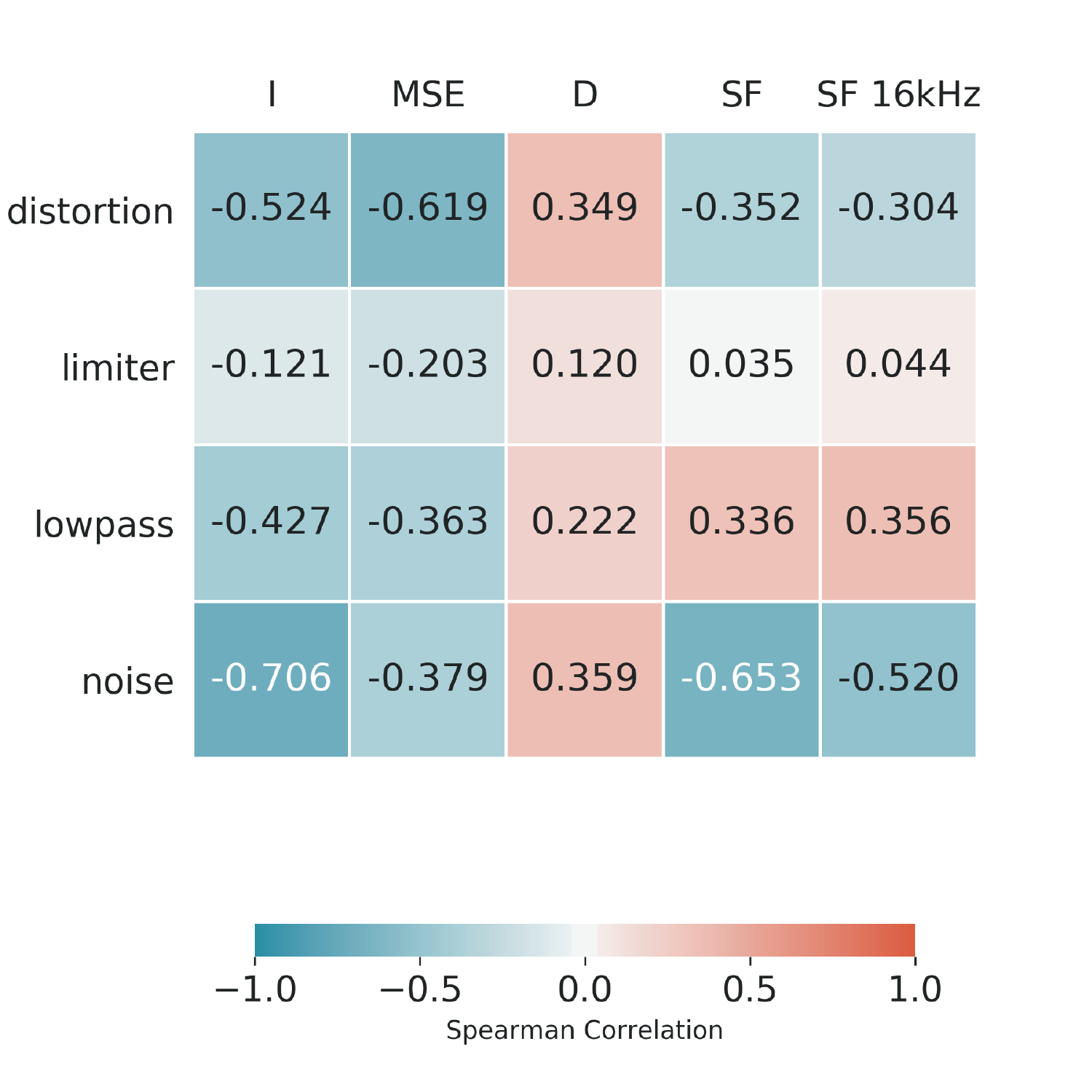}
        \subcaption{\label{fig:correlations:type}}
    \end{subfigure}
    \begin{subfigure}[b]{.32\linewidth}
        \includegraphics[width=\linewidth]{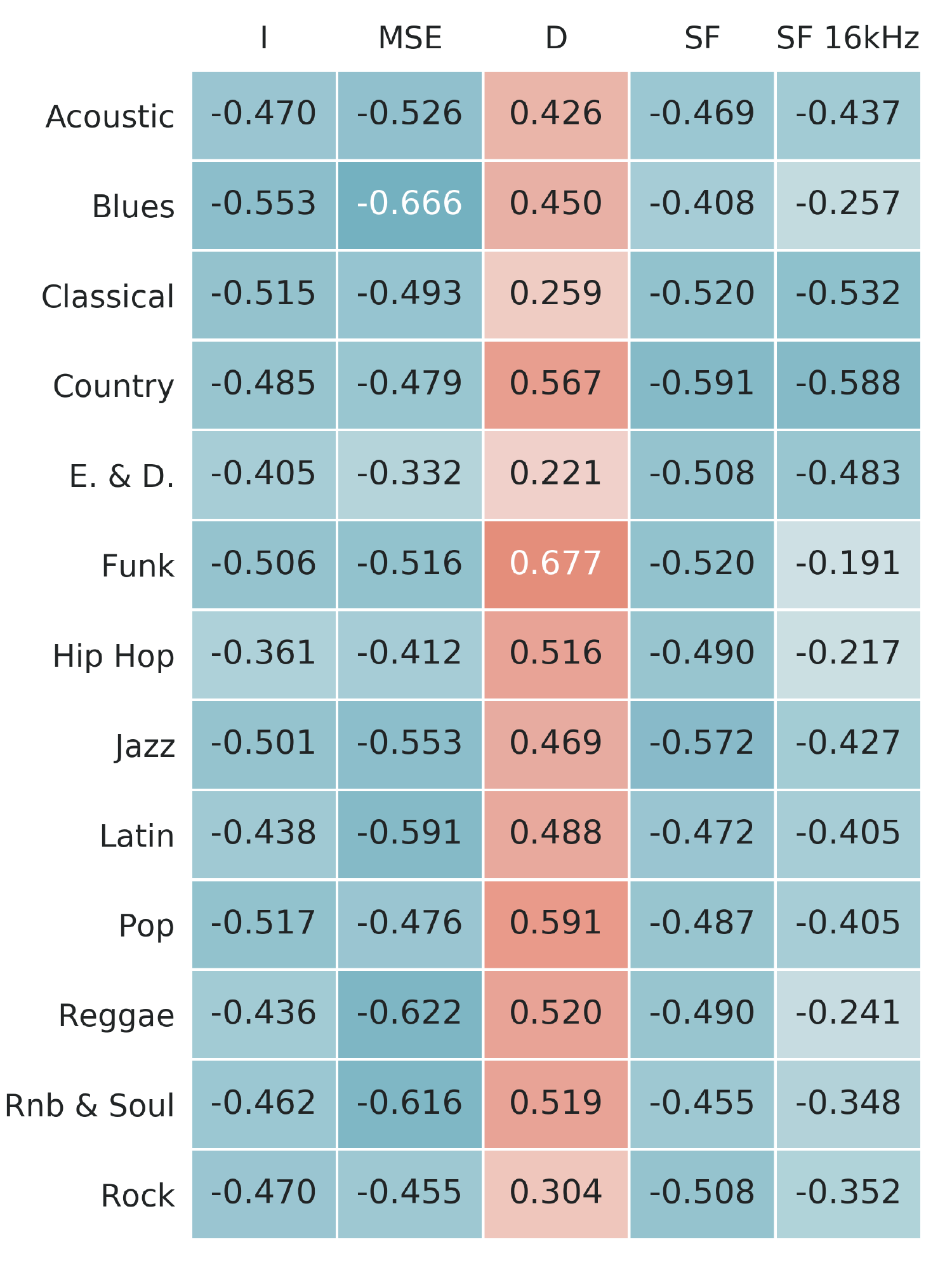}
        \subcaption{\label{fig:correlations:genre}}
    \end{subfigure}
\caption{\subref{fig:correlations:pairwise} Pairwise correlation between measures, \subref{fig:correlations:type} Correlation to median human rating by degradation type, \subref{fig:correlations:genre} Correlation to median human rating by genre. \textit{R} median human rating, \textit{I} intensity of degradation, \textit{\gls{mse}} between original and degraded clip, \textit{D} (ours) discriminator score, \textit{\acrshort{sf}} \acrlong{sf} at 48kHz and \textit{\acrshort{sf} 16kHz} \acrlong{sf} at 16kHz. Of particular interest is the sign change in \subref{fig:correlations:type} for \textit{\acrshort{sf}} and \textit{\acrshort{sf} 16kHz}. Furthermore, for some genres \textit{D} is significantly more correlated to the median rating than other measures, including \textit{\gls{mse}} and \textit{I}.}
\label{fig:correlations}
\end{figure*}
Since we are not familiar with any method for perceptual quality scoring without references we choose a number of well known measures to compare with \textit{D}. The two first are the known degradation intensity and a reference based metric, and thus do not comprise fair comparisons, yet help put the results in context. The selected measures are:
\begin{itemize}
    \item \textbf{I} Intensity of the degradation
    \item \textbf{\gls{mse}} Mean Squared Error between the original waveform and the final, possibly degraded waveform
    \item \textbf{\acrshort{sf}} \Acrlong{sf} of the audio at the original 48kHz / 24bit stereo content averaged over the entire clip
    \item \textbf{\acrshort{sf} 16kHz} \Acrlong{sf} of the audio downsampled to 16kHz / 16bit mono
\end{itemize}
\Gls{sf}, with implementation by \citeauthor{brian_mcfee_2019_2564164} \shortcite{brian_mcfee_2019_2564164}, was chosen since it has been designed to detect noise, and would form an interesting point of comparison due to the inclusion of noise in the degradations used. To illustrate what is possible at the sample rate available to our method we also evaluate \gls{sf} on the same 16kHz downsampled version of the segments consumed by the \gls{gan}. All these measures are shown by their correlation to each other and to the human rating broken down by different subsets in Figure \ref{fig:correlations}. Note that the listed measures are expected to have negative correlation with the rating, whereas our method (D) is expected to have a positive correlation, whereby the most relevant comparison is by magnitude.

The measure with the strongest absolute correlation to the human rating is \textit{\gls{mse}} at $-0.510$ (Figure \ref{fig:correlations:pairwise}). Our method \textit{D} is close in magnitude ($0.426$) and performs significantly better than \gls{sf} ($-0.345$) when using the same fidelity content, though slightly lower than \gls{sf} at the full 48 kHz / 24bit ($-0.473$). Despite this, \gls{sf} is not a generally useful predictor of human rating, as can be seen by the sign changes for both versions across different types of degradation (Figure \ref{fig:correlations:type}). Our method \textit{D} on the other hand maintains a monotonic correlation with human rating. Surprisingly, when broken down by genres (Figure \ref{fig:correlations:genre}), our method \textit{D} outperforms all other measures, including the parameter of the generating process \textit{I} and the reference based \textit{\gls{mse}}, on the genres \textit{Funk}, \textit{Hip-Hop} and \textit{Pop}.

\subsection{Effect of degradation intensity}
By studying column \textit{I} of Figure \ref{fig:correlations:type} the effect of the intensity for each degradation may be studied. It is clear that adding noise is by far the strongest detriment to perceived quality ($\rho_s(R, I) = -0.706$), whereas the limiter barely produces a significant effect on the rating ($\rho_s(R, I) = -0.121$). In Figure \ref{fig:correlations:genre} we see that the lowest rank correlation between intensity and rating across genres is for \textit{Hip Hop} ($-0.361$), which is likely due to the genre regularly incorporating some of the chosen degradations as sound effects.

\subsection{Generation}
\begin{figure}[ht]
\includegraphics[width=\columnwidth]{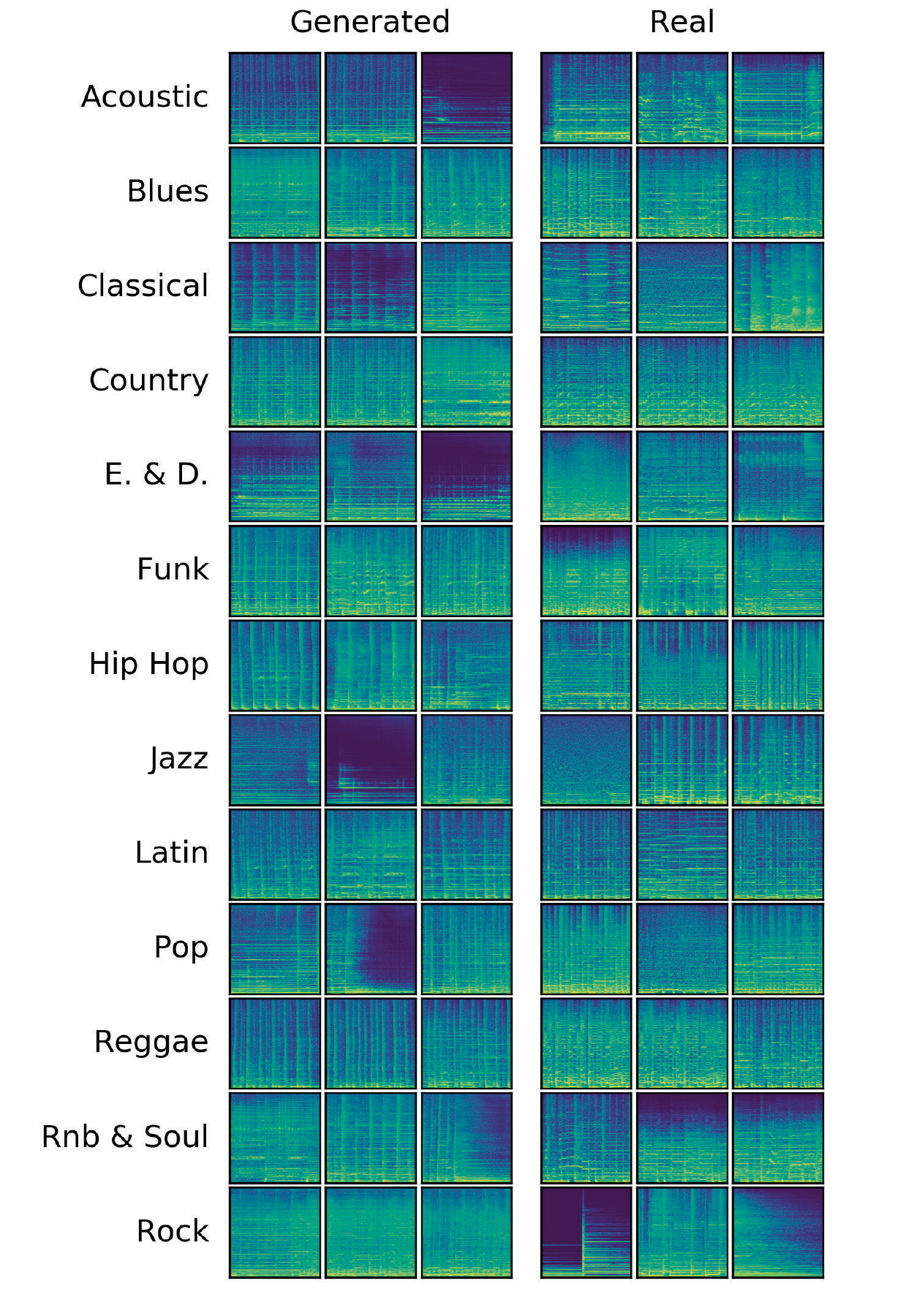}
\caption{\label{fig:spectrograms_by_genre}Generated and real log-scaled Mel spectrograms from each genre in the dataset.}
\end{figure}
As a final verification that the \gls{gan} does learn to model the distribution of mel spectrograms we qualitatively study a set of generated samples. A random sample of generated mel spectrograms is shown in Figure \ref{fig:generated_mel_spectrograms}. This shows that generated samples often contain clearly defined harmonics and is able to capture behaviors like clear attack, decay, sustain and release phases. Furthermore, we see expected differences between genres such as stronger harmonics in classical music whereas rock appears more distorted.

A larger sample of generated and real mel spectrograms is shown in Figure \ref{fig:spectrograms_by_genre}. By comparing the generated samples against the real we see several important differences. Most noticeable is that the real mel spectrograms tend to have stronger and more synchronized harmonics and larger variation among samples. Two common failures of the generated spectrograms are their unsynchronized attack phases and seemingly missing parts of spectrograms.

\section{Conclusion and discussion}
We have shown that a GAN discriminator is indeed predictive of the perceived quality of music. The discriminator score has a significant correlation with human perceived quality for the data presented. Compared to some constructed measures it performs favorably, showing a slightly weaker correlation than measures with reference or ground truth knowledge. The \acrlong{sf} at the full 48kHz / 24 bit material does show a correlation to the human rating of larger magnitude due to the strong effect of noise on the quality. It is for that reason, however, not generally applicable across different types of degradations. The discriminator score is shown to have a notably strong correlation with perceived quality for certain genres, including \textit{Hip-Hop}. This is of particular interest since \textit{Hip-Hop} is among the most challenging genres, seen by the weak correlation of degradation intensities and other measures to the perceived quality of content.

Interestingly, the GAN discriminator is able to perform this task without access to any type of degradation during training and without a reference at test time, making the method attractive to use. Though we do not discount the possibility that training discriminative models on annotated datasets might be fruitful, defining a broad range of negative examples (i.e. low quality musical audio) requires extensive domain knowledge of music production, composition and acoustics. In our work we are circumventing this by only modeling positive examples of high quality musical audio. This also means adapting it to new domains like other genres becomes simple, and requires no fundamental exploration of the applicable types of degradation.

\subsection{Suggested directions of future work}
The advantages of this method and these first positive results warrant further work into audio perception through generative modelling. Therefore, as a final remark, we would like to suggest a few directions for future work to further explore this method.

\begin{itemize}
    \item The human rated data quality should be improved. It would be interesting to not only increase the magnitude of the crowdsourced listening, but those findings should also in the future be expanded to include controlled trials to improve and verify the quality of the data.
    \item Improving the performance of the GAN. As we show in the results the generated mel spectrograms do exhibit certain convincing features yet show plenty of room for improvement. In particular, increasing the stability by a larger batch size during training such as in~\cite{brock2018large} would be a readily available method for improving the GAN.
    \item The discriminator's ability to perceive quality could be related to its importance in performing anomaly detection. As there are multiple methods of performing anomaly detection using GANs~\cite{anogan} it would be interesting to compare such methods to the one presented here.
    \item The discriminator score's correlation with human rating should be benchmarked against a more perceptually accurate metric such as PEAQ instead of \gls{mse}.
\end{itemize}

\bibliography{references.bib}
\bibliographystyle{aaai}
\end{document}